\documentclass[twocolumn,showpacs,preprintnumbers,amsmath,amssymb]{revtex4}

\usepackage{graphicx}
\usepackage{dcolumn}
\usepackage{bm}

\begin{document}

\title{Antiferromagnetic vs ferromagnetic interactions and spin-glass-like
behavior in ruthenates}
\author{P.Ravindran}
\email{ponniah.ravindran@kjemi.uio.no}
\homepage{http://folk.uio.no/ravi}
\author{ R.Vidya} 
\author{ P.Vajeeston}
\author{ A.Kjekshus} 
\author{H.Fjellv{\aa}g}
\affiliation{Department of Chemistry, University of Oslo, Box 1033,
Blindern, N-0315, Oslo, Norway}
\author{B.C. Hauback}
\affiliation{Institute of Energy Technology, P.O.Box 40, Kjeller, N-2007, Norway }
\date {\today}
\begin{abstract}
We have made a series of gradient-corrected relativistic full-potential density-functional
calculation for Ca-substituted and hole-doped SrRuO$_3$
in para, ferro, and $A$-, $C$-, and $G$-type antiferromagnetic states. 
Magnetic phase-diagram data for Sr$_{1-x}$Ca$_x$RuO$_3$
at 0~K are presented.
Neutron diffraction measurement combined with
total energy calculations show that spin-glass behavior with short-range
antiferromagnetic interactions rules in CaRuO$_3$.
The substitution of Sr by Ca in SrRuO$_3$ decreases the ferromagnetic
interaction and enhances the $G$-type antiferromagnetic interaction; the
$G$-AF state is found to stabilize around  
$x$ = 0.75 consistent with experimental observations.
Inclusion of spin-orbit coupling is
found to be important in order to arrive at the correct magnetic ground state in ruthenates.
\end{abstract}
\maketitle
Ever since unconventional superconductivity was observed in 
Sr$_2$RuO$_4$ \cite{maeno95}, ruthenates have attracted much interest. 
Substitution \cite{nakatsuji00} of the smaller Ca for Sr in Sr$_2$RuO$_4$ leads to 
an antiferromagnetic (AF) Mott insulator with a staggered moment of $S$ = 1. 
Coexistence of ferro- and antiferromagnetic fluctuations in Ca$_2$RuO$_4$ and
competition between $p$- and $d$-wave superconductivity in Sr$_2$RuO$_4$ have
also been reported \cite{mazin99}.
Recently 
coexistence of magnetism and superconductivity is 
discovered \cite{bauernfeind95} in 
ruthenium-based layered cuprates such as 
$R$RuSr$_2$Cu$_2$O$_8$ ($R$ = Eu, Gd).  Lee {\em et al.} \cite{lee01} reported
the existence of a pseudogap (PG) in BaRuO$_3$ reminiscent of the 
PG in high-$T_c$ superconductors. 
As in $f$-electron-based intermetallics, 
non-Fermi-liquid (NFL) behavior has recently been observed in La$_4$Ru$_6$O$_{19}$ \cite{khalifah01}. 
Critical magnetic fluctuations associated with metamagnetism is reported for 
Sr$_3$Ru$_2$O$_7$ \cite{perry01} and a metallic AF phase with temperature induced insulator-to-metal 
transition is found for Ca$_3$Ru$_2$O$_7$ \cite{cao971}.
\par 
Poorly metallic NFL-behaving SrRuO$_3$ is the only known ferromagnetic (F; $T_C$ = 160\,K) 
4$d$ transition-metal oxide \cite{callaghan66,kostic98}.
CaRuO$_3$ is also metallic, but experimental and theoretical studies conclude contradictory regarding the nature
of the magnetic ground state (e.g., AF \cite{callaghan66,longo68,martinez95}, 
nearly F \cite{kiyama99}, exchange enhanced
paramagnetic (P) \cite{yoshimura99}, Curie-Weiss P \cite{gibb73}, 
verge of F instability \cite{mazin97},
spin glass (SG) \cite{felner00} etc.), and this controversy is not settled yet.
Some of these reports indicate a lack of long-range magnetic order whereas others
suggest evidence for an AF ground state with
a N\'{e}el temperature ($T_{N}$) of $\sim$110\,K \cite{longo68}.
The striking difference in the magnetic properties of SrRuO$_3$ and CaRuO$_3$ makes
magnetic phase diagram studies on Sr$_{1-x}$Ca$_x$RuO$_3$ quite interesting.
\par
CaRuO$_3$ and
SrRuO$_3$ are isostructural and isoelectronic, differing structurally 
only in the degree of the small orthorhombic distortion (lattice constants
within 2$\%$). Hence, insight into the magnetic properties
of these compounds is expected to increase the general knowledge on magnetic phenomena
in perovskite oxides, and ruthenates in particular.
The density-functional calculations \cite{mazin97,santi97} for CaRuO$_3$ which conclude with
``verge of F instability'' have the weakness that AF interactions were not considered.
In this letter, we present the intriguing result that {\it SG behavior
with short-range AF interaction occurs in metallic CaRuO$_3$}.
\par
The present full-potential linear-muffin-tin orbital \cite{wills} (FPLMTO) calculations 
are all-electron, and no shape approximation to the charge density or
potential has been used. 
The basis functions, charge density and potential were expanded in spherical harmonic
series inside the muffin-tin spheres and in Fourier series in the interstitial regions.
The calculations are based on the generalized-gradient-corrected (GGA) density-functional
theory as proposed by Perdew {\em et al.} \cite{pw96}. 
Spin-orbit (SO) terms
are included directly in the Hamiltonian matrix elements
for the part inside the muffin-tin spheres.
The basis set contained semi-core $4p$ and valence $5s$,
$5p$, and $4d$ states for Sr, $4s$, $4p$, and $3d$ states for Ca, $5s$, $5p$, and $4d$
for Ru, $2s$, $2p$, and $3d$ states for O, and $4s$, $4p$, and $3d$ states for K.
All orbitals were contained in the same energy panel.
The self consistency was obtained with 284~{\bf k}
points in the irreducible part of the first Brillouin zone for the orthorhombic structure
and the same density of {\bf k} points
were used for the cubic structure as well as for the supercells.
Structural parameters for Sr$_{1-x}$Ca$_x$RuO$_3$ are taken from 
the X-ray diffraction data reported by Kobayashi {\em et al.} \cite{kobayashi94}.
Details about the experimental reexamination of CaRuO$_3$ will be presented
in the full account of our findings \cite{vidya}.
\par
There are three possible magnetic arrangements
according to the interplane
and intraplane couplings in perovskite oxides. 
(i) With interplane AF coupling and intraplane F coupling
the $A$-AF structure arises. (ii) The opposite arrangement with interplane F coupling and
intraplane AF coupling is called $C$-AF
structure.
(iii) If both the inter- and intraplane couplings are AF the $G$-AF
structure arises. In the $G$-type AF lattice, each Ru atom will be surrounded by six Ru neighbors whose
moments are antiparallel to that of the chosen reference atom. The details about the
calculations on magnetic and excited state properties can be found elsewhere \cite{ravi_lamn}.
We have placed the magnetic moment direction along [001] in all calculations to 
comply with the 
out-of-plane easy axis observed for Sr$_{1-x}$Ca$_x$RuO$_3$ \cite{cao97}.
The energy differences between the various magnetic configurations are very small and
extreme computational accuracy is therefore needed.
The theoretical method used in the present study is capable of
reproducing energy differences of the order of $\mu$eV \cite{mae}.
\par
The calculated total energy for CaRuO$_3$ from the GGA calculation with (GGA+SO)
and without SO coupling 
are given in Table~\ref{table:ene} relative to the corresponding
ground state.  From this table it
is clear that the $G$-AF phase is the ground state for CaRuO$_3$. The total energy differences
between the different AF phases are as already mentioned very small. If we disregard the 
orthorhombic distortion,
the $F$ phase is seen to have the lowest energy indicating that 
strong magneto-elastic couplings are present in CaRuO$_3$. We also see large energy gains
when we include the orthorhombic distortion in the calculation.
The F phase is found to be the ground state with 7\,meV/f.u. lower energy
than the $G$-AF phase when SO interactions are neglected.  
Hence, {\it SO coupling appears to play an important role in
deciding the magnetic properties of ruthenates.}
\begin{table}
\caption{\label{table:ene}
Total energy (relative to the lowest energy state; in meV/f.u.) for CaRuO$_3$
in P, F, and $A$-, $C$-, and $G$-AF states with orthorhombic and undistorted cubic (designated
cubic) structures.
}
\begin{ruledtabular}
\begin{tabular}{lccccl}
Method                &   P   &  F    &     $A$-AF  &  $C$-AF   &  $G$-AF  \\
\hline
GGA+SO         &  26  &  19.8  &  6     &  4  & 0      \\
GGA+SO (cubic)  &  1307 &  1249 &  1292        &  1299 & 1303  \\
GGA         &  38    & 0    &  9     & 10    & 7  \\
\end{tabular}
\end{ruledtabular}
\end{table}
\par
The calculated magnetic moment at the Ru site in SrRuO$_3$ is 0.92\,$\mu_{B}$/atom
is in good agreement with 0.85\,$\mu_{B}$/atom
obtained from magnetization measurements \cite{callaghan66}.
Goodenough \cite{goodenough76} has argued that, in general, a spontaneous change between F
and AF states are expected when $\mu_{F}\approx\mu_{AF}$,
the stable magnetic phase being the one with the highest
atomic moment. Our findings are consistent with this viewpoint in
that the larger magnetic moments are found for the stable
magnetic configuration of Sr$_{1-x}$Ca$_x$RuO$_3$ (viz., for $x <$ 0.724 the F phase
has highest magnetic moment and for $x >$ 0.724 $G$-AF has the highest moment).
The hyperfine field at the Ru site in $G$-AF state CaRuO$_3$ is
263\,kOe/$\mu_B$ in agreement with the experimental
value of 222 $\pm$ 50\,kOe/$\mu_B$ obtained from Knight shift measurements \cite{mukuda99}.
\par
\begin{figure}
\includegraphics[scale=0.45]{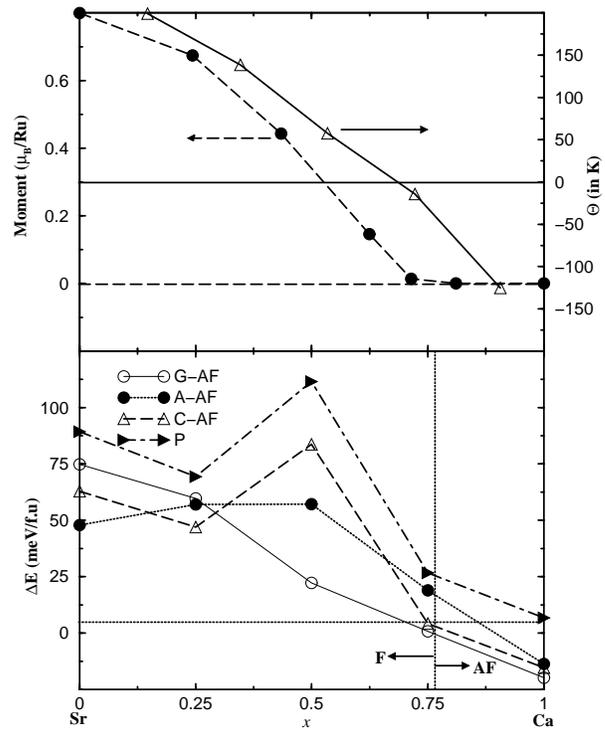}
\caption{\label{fig:phase} Magnetic phase diagram data for Sr$_{1-x}$Ca$_x$RuO$_3$. Energies
are given with respect to the total energy for the F
phase.}
\end{figure}

The experimentally found spontaneous magnetic moment \cite{kiyama99} and 
Weiss temperature \cite{fukunaga94} for Sr$_{1-x}$Ca$_x$RuO$_3$ are given in Fig.~\ref{fig:phase}
along with the calculated total energy difference relative to that of the F phase
for the various magnetic configurations. 

An increase in $x$ decreases the F interaction and increases the G-AF
interaction. 
Because of the smaller ionic radius of Ca compared with Sr (0.99 vs
1.18\,{\AA}), tilting of the octahedra occur when
one substitute Ca for Sr in SrRuO$_3$ which in-turn affect the magnetic properties. 
Fig.~\ref{fig:phase} shows that the F to $G$-AF transition takes place
around $x$ = 0.75 both experimentally 
and theoretically. {\it This confirms the presence of AF
interactions in CaRuO$_3$.} One of the interesting aspects of Fig.~\ref{fig:phase}
is that at $x$ = 0.5, there is large gain in the total energy of F phase compared
with P, $A$- and $C$-AF phases. DOS analyses show that the F phase of Sr$_{0.5}$Ca$_{0.5}$RuO$_3$
is nearly half-metallic with a gap of 0.73~eV in the majority spin channel
which gives extra contribution to the stability.
Moreover, $E_{F}$ is at peaks in the DOS curves of the P, $A$- and $C$-AF phases 
(which further contribute to the enhanced energy difference relative to the F state).

\begin{figure}
\includegraphics[scale=0.5]{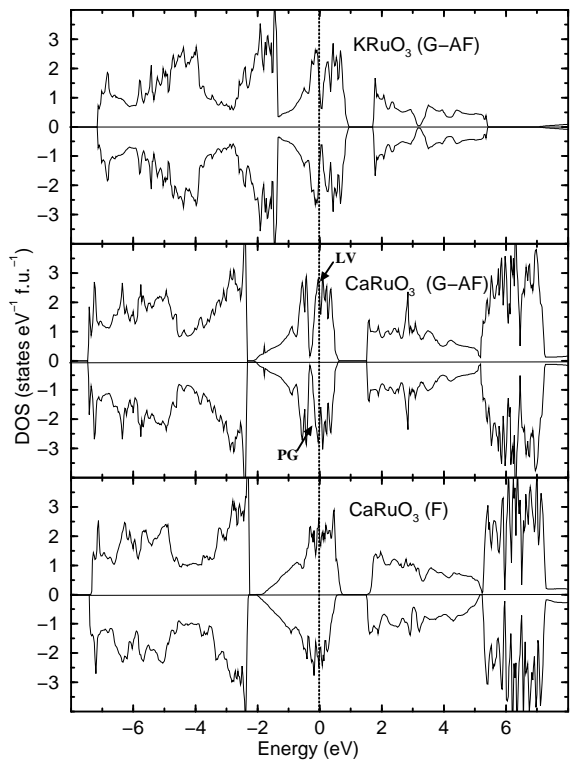}
\caption{\label{fig:dos} Total DOS for CaRuO$_3$ in F and $G$-AF states
and hypothetical KRuO$_3$ in $G$-AF state. Fermi level is set to zero.}
\end{figure}

\par
\begin{figure}
\includegraphics[scale=0.5]{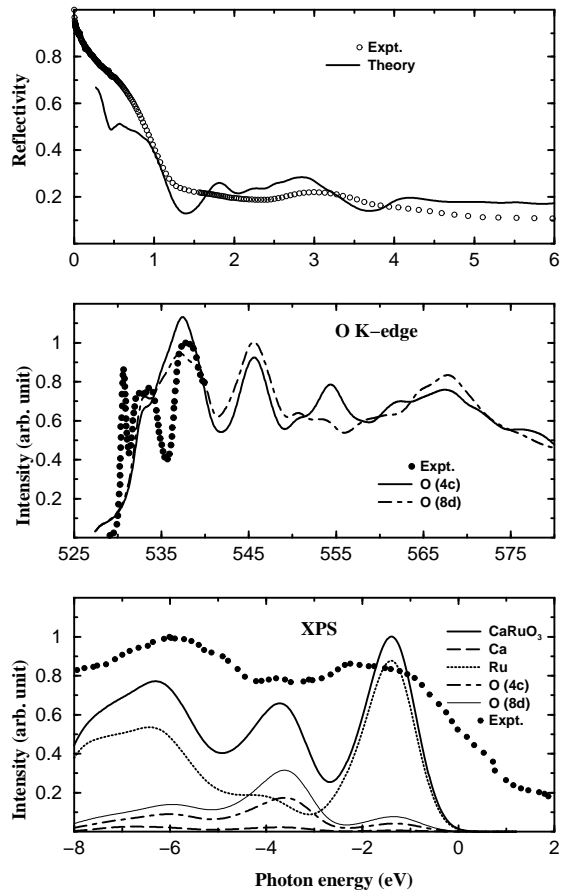}
\caption{\label{fig:spectra}Calculated valence band XPS, O $K$-edge XANES, 
and optical reflectivity spectrum for $G$-AF state CaRuO$_3$. Experimental XPS, XANES and reflectivity
spectra are taken from Refs.~[\protect\onlinecite{rama01,lee01x}].}
\end{figure}

The lower total energy for the $G$-AF state for $x$ $>$ 0.724 compared with the 
other magnetic configurations can be understood 
as follows. 
The spin-projected total DOS curve for CaRuO$_3$ in the
F and $G$-AF phases are shown in the lower and middle panel of 
Fig.~\ref{fig:dos}. In the F case $E_{F}$ is at a peak-like feature
in the DOS curve. This infers an instability to the lattice. However,
for the $G$-AF phase a deep valley-like  (PG-like) feature appears in
the DOS curve. This will give an extra contribution
to lattice stability \cite{ravi97} and hence CaRuO$_3$ is expected to stabilize in
the $G$-AF phase. The calculated excited-state properties such as reflectivity,
O $K$-edge spectra and XPS spectra for CaRuO$_3$ in the $G$-AF phase are found to be
in good agreement with experimental spectra (see Fig.~\ref{fig:spectra}), indicating 
that the ground state is correctly assigned.{\it The creation of a PG-like feature on
introduction of G-AF ordering gives an extra contribution to the band energy term of the
total energy (66\,meV/f.u.) which stabilizes the G-AF phase over the F phase.}
However, our neutron powder diffraction diagrams at 298 and 8~K 
show no sign of extra Bragg reflections 
at 8\,K, which should have been the proof of long-range AF ordering. (A full account of
the neutron diffraction findings will be given in the forthcoming article \cite{vidya}.)
Hence, the magnetic exchange interactions must have only short-range influence.
Closer examination of the DOS
curve for the $G$-AF phase shows that $E_{F}$ is located on a shoulder of a 
peak-like feature (an unfavorable condition for stability) 
with a local valley (LV, see Fig.\,\ref{fig:dos}) situated just 40\,meV above $E_{F}$. 
This may explain why there is no established long-range magnetic ordering in this material.
Recent magnetization measurements \cite{felner00}
show distinctions between in the zero-field-cooled and field-cooled characteristics
that may indicate SG behavior. So, we conclude that {\it CaRuO$_3$ exhibits
short-range AF ordering which manifests itself in SG behavior at low temperature}.
\par
Rigid-band-filling analysis show
that an addition of 0.136 electrons/f.u. will bring 
$E_{F}$ to LV with the expected effect of long-range $G$-AF ordering. 
In fact, long-range magnetic
ordering has been observed for CaRu$_{1-x}$Rh$_x$O$_3$ with $x > 0.015$
where Rh acts as an electron donor for CaRuO$_3$ \cite{cao973}.
Rigid-band-filling analysis also show that removal of
exactly one electron from CaRuO$_3$ will bring $E_{F}$
to a PG. In order to test this possibility we have made
additional calculations for hypothetical KRuO$_3$ with the structural
parameters for CaRuO$_3$. Consistent with the rigid-band-filling findings $E_{F}$
is found at PG for the $G$-AF phase of hypothetical KRuO$_3$ (see upper panel
of Fig.~\ref{fig:dos}). When 
$E_{F}$ is at PG, all bonding states would be filled and all antibonding
states empty, implying extra contributions to stability. 
Total energy 
calculations show that the $G$-AF phase is 21.7\,meV/f.u. lower in energy than
the F phase for hypothetical KRuO$_3$, viz. a larger difference than that for CaRuO$_3$ indicating that
one really should expect $G$-AF long-range ordering in KRuO$_3$ at low temperature. 
Further the $G$-AF phase is
38.6 meV/f.u. lower in energy than the P phase for hypothetical KRuO$_3$ indicating
that $T_{N}$ should be reasonably high.
It is worth to recall that
according to Goodenough \cite{goodenough63} the sign of the transfer integral
for 180$^{o}$ cation-anion-cation superexchange interactions between 
octahedral-site cations are predicted to give AF for Ru$^{5+}$-O-Ru$^{5+}$.
Ru$^{5+}$ ions with enhanced AF interaction has recently been established
for CaRu$_{0.95}$Cu$_{0.05}$O$_3$ \cite{bradaric01}.
Moreover, the change in the oxidation state of Sr$_3$Ru$_2$O$_7$ from Ru$^{4+}$ 
to Ru$^{5+}$ by fluorine addition stabilizes $G$-AF ordering \cite{li00}.
Oxygen nonstochiometry and hole doping can accordingly bring $E_{F}$ toward
PG and stabilize the $G$-AF ordering. This may explain why
long-range AF ordering has been reported for some experimentally studied CaRuO$_3$
samples \cite{longo68} and Na doped CaRuO$_3$ \cite{shepard96}.
\par
In the SG state the moments are arranged in a certain equilibrium oriented pattern,
but without long-range order. 
Some characteristics of SG materials are: (i) Existence of local moment
as we have found with spin-polarized calculations. 
(ii) No magnetic Bragg scattering at low temperature 
as our neutron diffraction study show. (iii) A history-dependent magnetic response
as the recent magnetization measurements \cite{felner00} have shown.
(iv) Both long-range and short-range terms in the
Hamiltonian \cite{chowdhury86}. In accordance with our calculations 
we believe that short-range nearest-neighbor AF interactions are dominant in CaRuO$_3$.  
\par
Two important ingredients necessary to produce SG behavior are
frustration and partial randomness of the interaction between the magnetic
moments \cite{fischer91}. Owing to the small total energy differences between the
AF and F phases
of CaRuO$_3$ (see Table\,\ref{table:ene} and Fig.~\ref{fig:phase}),
there will be a competition among the different AF and F interactions, in the sense
that no single configuration of the the moments is uniquely favored by all 
interactions ($viz$. frustration). 
In other words, the AF moments can arrange themselves randomly in small domain-like 
regions with minimal loss in energy. 
The Ru ions which are responsible for the
magnetic properties of CaRuO$_3$ are in a strictly periodic order. Hence, structural
disorder can not be responsible for the SG behavior. The SG state of CaRuO$_3$ is characterized by a
predominant AF situation. 
AF systems with SG-like transition have been observed in ruthenates with
pyrochlore-like structures such as $R_2$Ru$_2$O$_7$ for
which recent experimental results indicate that the atomic arrangement does not
participate in the P-to-SG transition which is solely associated
with the Ru moments \cite{ito01}.
{\it The small energy differences between the different AF states bring disordering of the
moments and hence SG behavior to CaRuO$_3$ at low temperature.} 
In fact, SG behavior has been found for CaRuO$_3$ doped with small amounts 
of Sn \cite{cao961} or Rh \cite{cao973}. 
The calculated excited state properties for CaRuO$_3$ in the $G$-AF state are
in good agreement with the experimental measurements indicating that the electronic
structure does not differ significantly between the SG and $G$-AF phases.
\par
We conclude that CaRuO$_3$ exhibits short-range AF interaction with 
SG behavior. The hitherto hypothetical KRuO$_3$  
is expected to exhibit long-range $G$-AF ordering.
We have demonstrated that the relative strengths of
the F and AF exchange interactions can
be varied by varying the Ca content of Sr$_{1-x}$Ca$_x$RuO$_3$. 
\par
PR is grateful to 
the Research Council of Norway for financial support. Part of these
calculations were carried out on the Norwegian supercomputer facilities.

\end{document}